\documentstyle[prl,aps,psfig,subfigure]{revtex}
\def \greatersim{\mathrel{\lower1pt\hbox{$>\atop\raise2pt\hbox{$\sim$}$}}}
\begin{document}
\draft
\twocolumn[\hsize\textwidth\columnwidth\hsize\csname
@twocolumnfalse\endcsname
\title{Superfluid to solid crossover in a rotating Bose-Einstein condensed gas}
\author{David L. Feder$^{1,2}$ and Charles W. Clark$^2$}
\address{$^1$Institute for Physical Science and Technology, University of
Maryland, College Park, MD 20742}
\address{$^2$Electron and Optical Physics Division, National Institute of
Standards and Technology, Gaithersburg, MD 20899-8410}
\date{\today}
\maketitle
\begin{abstract}
The properties of a rotating Bose-Einstein condensate confined in a prolate
cylindrically symmetric trap are explored both analytically and numerically.
As the rotation frequency increases, an ever greater number of vortices are
energetically favored. Though the cloud anisotropy and moment of inertia
approach those of a classical fluid at high frequencies, the observed vortex
density is consistently lower than the solid-body estimate. Furthermore, the
vortices are found to arrange themselves in highly regular triangular arrays,
with little distortion even near the condensate surface. These results are 
shown to be a direct consequence of the inhomogeneous confining potential.
\end{abstract}
]

\narrowtext

One of the most striking properties of liquid $^4$He(II) is its ability to
mimic the behavior of a solid body when subjected to uniform rotation. Since
the superfluid velocity field ${\bf v}_s$ is irrotational
($\nabla\times{\bf v}_s=0$), the superfluid component of $^4$He(II) might be
expected to remain at rest while the normal component rotates with the
container. In fact, for sufficiently large values of the rotation frequency
$\Omega$, the entire fluid is found to rotate like a classical liquid at all
temperatures~\cite{Osborne}. The paradox may be resolved by assuming that
the
superfluid is threaded by quantized vortices. These are singularities of
${\bf v}_s$, around which the phase of the superfluid order parameter
increases by 2$\pi$.
Although the mechanisms for the spin-up of the superfluid are not fully
understood, at equilibrium the vortices must flow with the normal velocity
due to the the mutual friction between superfluid and normal
components~\cite{Khalatnikov}. In addition to considerable indirect evidence
for this hypothesis, small numbers of vortices have been imaged directly in
rotating superfluid $^4$He~\cite{Donnelly}.

In this work we show how the presence of quantized vortices can allow a
Bose-Einstein condensate (BEC) to mimic a classical fluid under rotation, as
has been suggested by recent experiments at JILA~\cite{Haljan}. In these
experiments, a trapped gas of ultracold $^{87}$Rb atoms is spun up, and then
cooled through the Bose-Einstein condensation transition. For small values
of
$\Omega$, the condensate density is found to assume its usual non-rotating
shape, while the thermal cloud bulges outward. This corroborates previous
evidence that the condensate behaves as an irrotational
superfluid~\cite{Marago,Onofrio}. The condensate density profile undergoes a
sudden change at a value of $\Omega$ that is comparable to the thermodynamic
critical frequency for the stability of a single vortex~\cite{Fetter}. As
$\Omega$ increases further, the shape of the condensate gradually approaches
that of the thermal cloud. This suggests that for any given value of
$\Omega$ and temperature, the condensate contains the appropriate number and
distribution of vortices for thermodynamic equilibrium. In contrast,
when no appreciable thermal fraction is present, higher rotation frequencies 
are generally required to nucleate vortices in 
BECs~\cite{Madison,Abo-Shaeer,Hodby}.

We present two key results, which also bear on issues raised by recent
experiments
at MIT~\cite{Abo-Shaeer}. First, the vortices are arranged in extremely
regular triangular arrays: even near the condensate surface,
little circular distortion~\cite{Campbell} is found. Second, the number of 
vortices is consistently lower than that required to ensure solid-body
rotation throughout the condensate.

To make explicit comparison with the recent JILA experiment, we consider
the case of $N=200,000$ atoms of $^{87}$Rb, confined in a cylindrically
symmetric trap with radial frequency $\omega_{\rho}/2\pi=8$~Hz, and
anisotropy $\lambda\equiv\omega_z/\omega_{\rho}={5\over 8}$. 
Unless stated
explicitly, our units of energy, angular frequency, length, and 
time are given by $\hbar\omega_{\rho}$, $\omega_{\rho}$, 
$d_{\rho}=\sqrt{\hbar/M\omega_{\rho}}\approx 3.845$~$\mu$m, and 
$\omega_{\rho}^{-1}$, respectively, where $M$ is the
atomic mass and $\hbar$ is Planck's constant $h$ divided by $2\pi$.
We work in a frame that rotates with angular frequency $\Omega$
about the $z$ axis. 
The time-dependent Gross-Pitaevskii (GP) equation~\cite{GP},
which governs the dynamics of the condensate wavefunction $\psi$ of a dilute
BEC at zero temperature, is then given by:

\begin{equation}
i\partial_t\psi({\bf r},t)
=\left[T+V_{\rm trap}+V_{\rm H}-\Omega L_z\right]\psi({\bf r},t),
\label{gp}
\end{equation}

\noindent with kinetic energy $T=-{\case1/2}\nabla^2$, 
trap potential $V_{\rm trap}={\case1/2}\left(\rho^2+\lambda^2z^2\right)$,
and angular momentum component
$L_z=i\left(y\partial_x-x\partial_y\right)$.
The effects of atomic interactions are included in the nonlinear term
$V_{\rm H}=4\pi\eta|\psi|^2$, $\eta=Na/d_{\rho}$,
where $a=5.29$~nm is the scattering length for $^{87}$Rb
collisions~\cite{Eite}.  We use the normalization condition
$\int d{\bf r}|\psi({\bf r},t)|^2=1$. In equilibrium in the rotating
frame, $\psi({\bf r},t)=e^{-i\mu t}\psi({\bf r})$, where $\mu$
is the chemical potential.

To estimate the properties of a rotating condensate, such as the
aspect ratio and the number of vortices, we consider two tractable cases: a
single vortex applicable for small $\Omega$, and multiple vortices relevant
to high $\Omega$ where the condensate is expected to behave essentially as a
rigid body.

With one vortex at the center of the trap,
$\psi = |\psi| e^{i\phi}$, where $\phi$ is the polar angle. In the large-$N$
or Thomas-Fermi (TF) limit, the condensate density is
$|\psi|^2=\left(\mu-{\rho^2\over 2}-{\lambda^2z^2\over 2}
-{1\over 2\rho^2}+\Omega\right)/4\pi\eta$ when that quantity
is positive, and is zero elsewhere.~\cite{Fetter}. 
The inner cutoff defines the vortex core
size or the healing length $\xi$; for $\mu\gg 1$, one obtains
$\xi\approx 1/\sqrt{2\mu}\sim 1/R_0$, where $R_0=(15\eta\lambda)^{1/5}$ is
the
TF radius along
$\hat{\rho}$ in the absence of a vortex. A straightforward calculation shows
that, for large $R_0$, the TF radius for an isolated vortex
is $R_{\rho}\approx R_0[1+(3/2R_0^4)\ln(2R_0/\xi)]$ and the condensate
aspect
ratio is $\lambda_{\rm TF}'\equiv R_\rho/R_z\approx\lambda[1+(1/2R_0^2)].$
Assuming this result depends only weakly on vortex position, and is additive 
with respect to the number of vortices $N_v$, then in explicit
units $\lambda_{\rm TF}'\approx\lambda[1+{1\over 2}N_v(d_{\rho}/R_0)^2]$ at
larger $\Omega$.

For large $N_v$, as we will show below, the condensate
rotates almost as a solid body, so the rotating-frame 
velocity operator ${\bf v}_r = -i \nabla - \Omega\hat{z}\times{\bf r}$
can be neglected.
Since $T-\Omega L_z
={1\over 2}{\bf v}_r ^2
-{1\over 2}\Omega^2\rho^2$,  rotation effectively
softens the radial potential,
$V_{\rm trap}\to{1\over 2}(1-\Omega^2)\rho^2+{1\over 2}\lambda^2z^2$. In
this
case, $R_{\rho}=R_0/(1-\Omega^2)^{3/10}$ and $\lambda_{\rm sb}'=\lambda/
\sqrt{1-\Omega^2}$. The number of vortices is the line integral of the phase
gradient around the cloud perimeter; assuming the solid-body value of the
tangential velocity $\Omega R_{\rho}$, then the areal vortex density is
$n_v=N_v/\pi R_{\rho}^2=\Omega/\pi$, and $N_v^{\rm sb}=\Omega R_0^2
/(1-\Omega^2)^{3/5}$.

If the vortices form a regular array at large $\Omega$, then the lattice 
constant $b$ should be comparable to the average separation between vortices
$n_v^{-1/2}\sim\sqrt{\pi/\Omega}$. For a triangular array centered at the 
origin~\cite{Tkachenko,Ho}, the vortices arrange themselves in concentric 
hexagonal rings labelled by ring index $r$, such that $N_v=1+3r(r+1)$. Assuming
the superfluid velocity exactly matches the solid-body value midway between
nearest-neighbor vortices (where the two nearest rotational fields exactly
cancel), then $N_v=\Omega b^2(r+{1\over 2})^2$ and $b\approx\sqrt{3/\Omega}$
for large $r$. Since $\Omega_{\max}=1$ in a harmonic trap, the smallest
vortex 
separation is $b_{\rm min}\approx\sqrt{3}d_{\rho}$ in explicit units. When
the vortex cores begin to overlap significantly ($b\sim\xi$), the system
might
undergo a phase transition, possibly into a state akin to a quantum Hall 
insulator~\cite{Ho}; since $\xi(\rho=0,z=0)=1/\sqrt{2\mu}
=1/R_0(1-\Omega^2)^{1/5}$ in the TF limit, the value of $\Omega$ for this to
occur must become extremely close to unity: $1-\Omega\sim R_0^{-5}$.

The stationary solutions of the GP equation in the rotating frame, defined
as local minima of the free energy $\langle E\rangle=\mu N-{1\over 2}
\langle V_H\rangle$, are found numerically by norm-preserving imaginary time
propagation using an adaptive stepsize Runge-Kutta integrator. The wavefunction
is solved on a three-dimensional Cartesian mesh within a discrete-variable
representation~\cite{Feder} based on Gauss-Hermite quadrature, and is assumed 
to be even under reflection in the $z=0$ plane. The initial state is taken to 
be the TF wavefunction with a phase
$\Phi(x,y)=\sum_{x_0,y_0}\tan^{-1}[(y-y_0)/(x-x_0)]$, where $(x_0,y_0)$ are 
vortex positions in a regular array centered at the origin. The GP equation
for a given value of $\Omega$ is propagated in imaginary time until the 
fluctuations in both $\mu$ and the norm become smaller than $10^{-11}$. The
condensate densities integrated down $\hat{x}$ and $\hat{z}$ are then fit to
a TF profile using a nonlinear least-squares analysis, where densities
lower than $0.1\%$ of the maximum value are discarded. For the vortex-free
condensate, this yields an aspect ratio of $0.645$, which is $3\%$ larger
than the TF value of ${5\over 8}$.

\begin{figure}[tb]
\centering
\subfigure{\psfig{figure=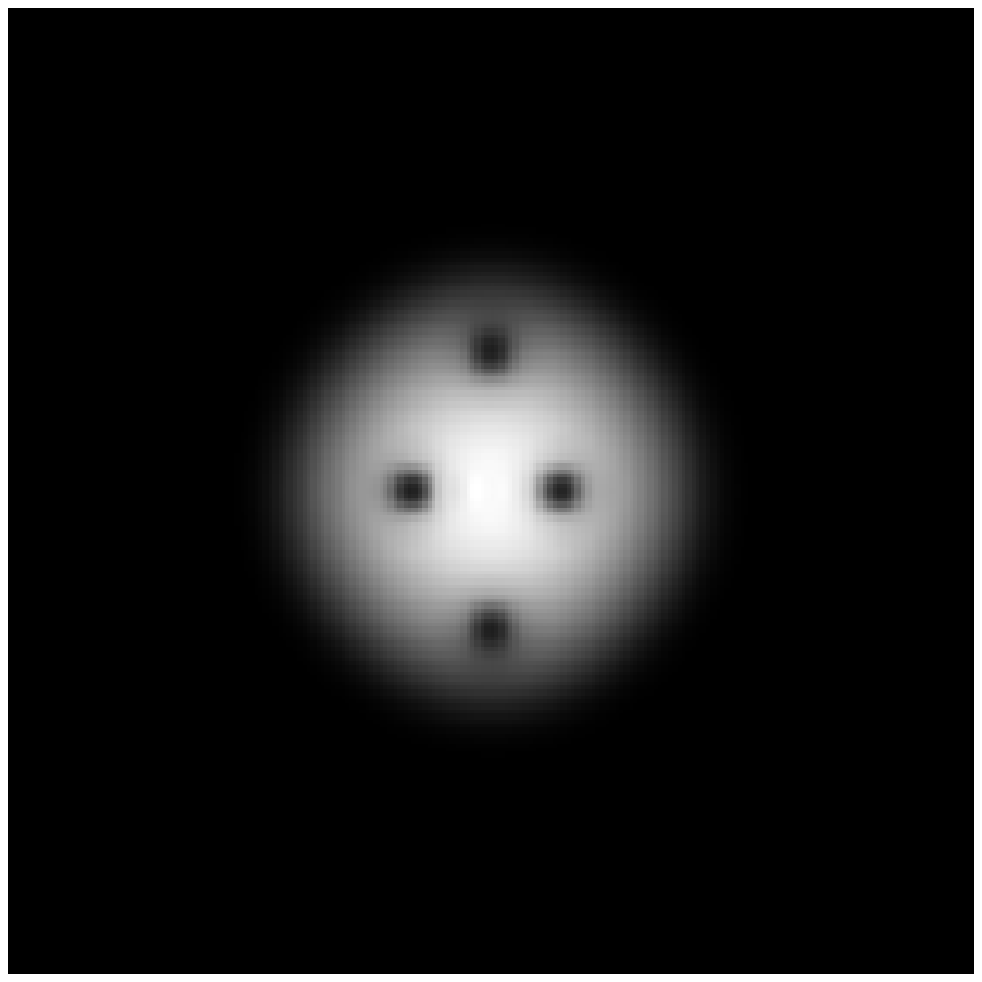,width=0.33\columnwidth,angle=0}}
\hspace{-0.2in}
\subfigure{\psfig{figure=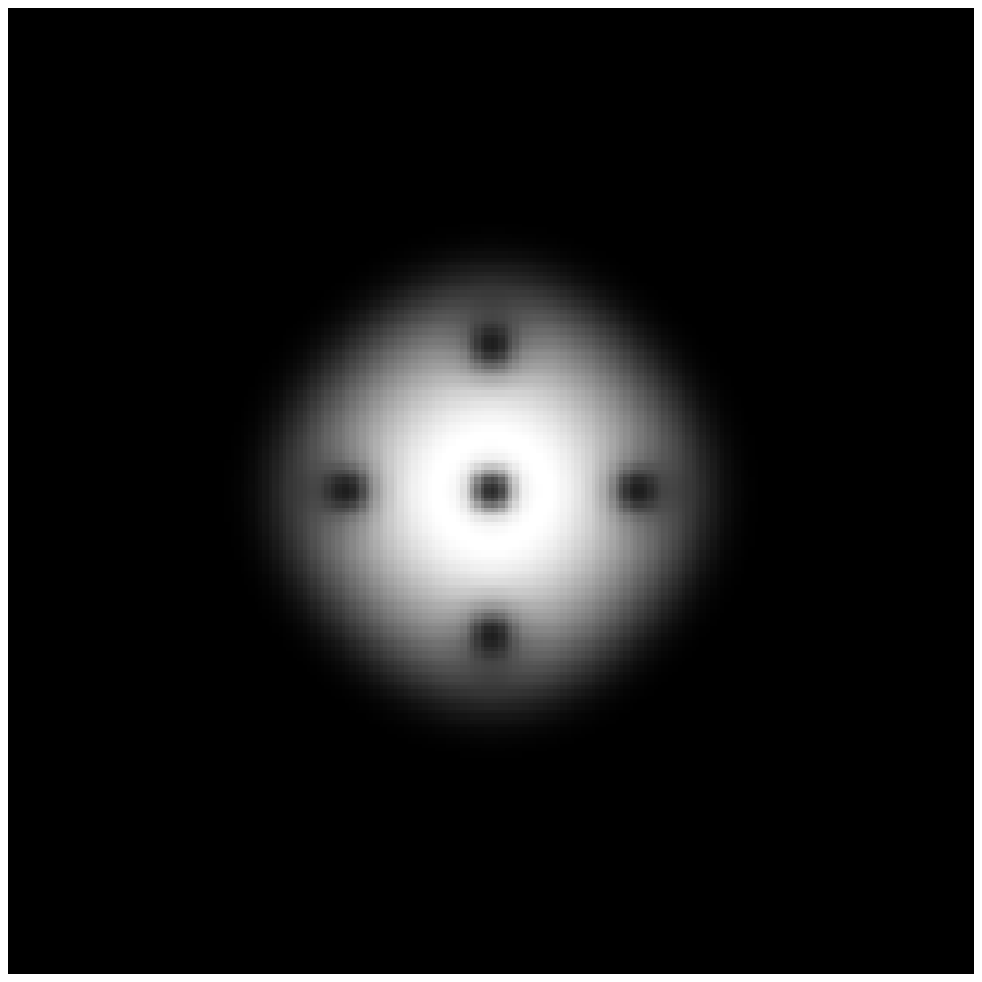,width=0.33\columnwidth,angle=0}}
\hspace{-0.2in}
\subfigure{\psfig{figure=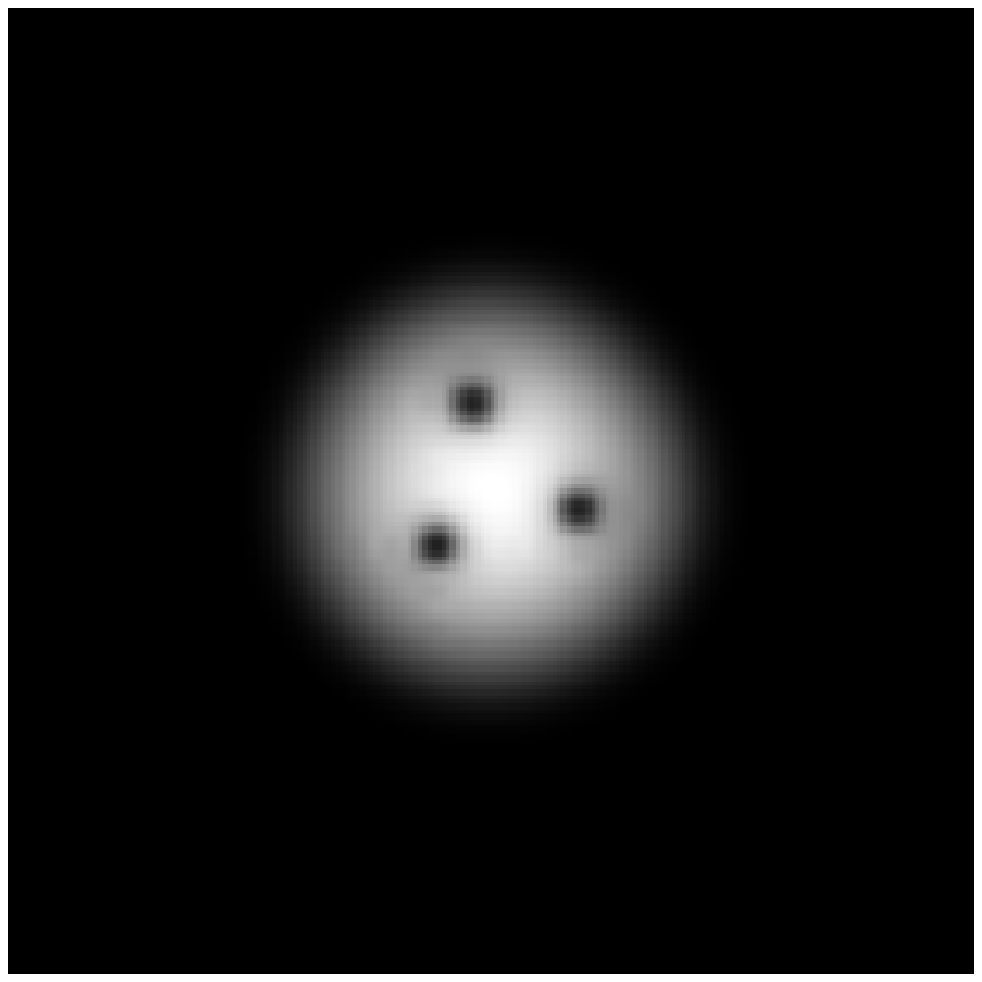,width=0.33\columnwidth,angle=0}}
\caption{Three stationary states are shown for $\Omega=0.45$. From left to 
right, the values $\{E/N, \mu, \langle L_z\rangle/N\}$ in scaled units are
$\{7.20, 11.28, 2.31\}$, $\{7.21, 11.29, 2.33\}$, and $\{7.24, 11.28,
2.19\}$,
respectively.}
\label{45fig}
\end{figure}

The resulting equilibrium configurations are sensitive to the initial vortex 
distributions. Fig.~\ref{45fig} shows three different solutions of the
GP equation~(\ref{gp}) for $\Omega=0.45$. These were obtained using seed
arrays with rhombohedral (left), square (center), and triangular (right) 
symmetries, respectively. Though observables such as the energy, angular
momentum, and cloud aspect ratio are all comparable, they each have
different vortex numbers and arrangements. Though a complete survey of possible
configurations is beyond the scope of the present work, for all cases
considered the initial rhombohedral vortex distribution is found to yield the 
final state with lowest energy; for larger $\Omega$ this symmetry gives rise 
to equilibrium arrays that are generally triangular (see below).

\begin{figure}[tb]
\psfig{figure=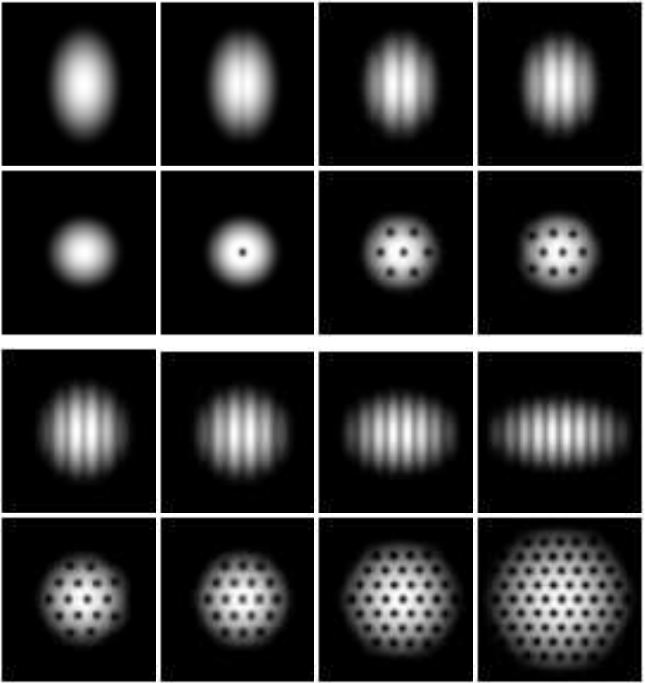,width=\columnwidth,angle=0}
\caption{Condensate densities integrated along $\hat{x}$ (upper row) and
$\hat{z}$ (lower row) are shown for $\Omega=0.25$, $0.35$, $0.55$, and $0.65$
(first data set, left to right) and $\tilde{\Omega}=0.75$, $0.80$, $0.90$,
and $0.95$ (second data set, left to right). Each frame is $20d_{\rho}\approx
77$~$\mu$m on a side.}
\label{pix}
\end{figure}

The central results of the present work are shown in Fig.~\ref{pix}, which 
depicts equilibrium solutions for $0.25\leq\Omega\leq 0.95$. A
single vortex at the origin has appeared by $\Omega=0.35$; the thermodynamic
critical frequency (the energy difference between states with zero and one
vortex, divided by $\hbar$) is $\Omega_c=0.30$. This value is slightly lower
than the experimental value $0.32<\Omega_c<0.38$; since
$\Omega_c\sim N^{-2/5}$, perhaps there are fewer atoms in the condensate at
vortex nucleation. 
(The dynamic critical frequency, at which the first 
collective mode becomes negative, is somewhat higher: $\Omega_{\nu}=0.46$).
With a vortex, the cloud aspect ratio changes to $\lambda=0.663$; using the 
fitted values for the nonrotating cloud $\lambda=0.645$ and $R_0=4.86$, the 
TF prediction is $\lambda_{\rm TF}'=0.659$. As $\Omega$ continues to increase,
so does the aspect ratio; the cloud becomes spherical for $0.75<\Omega<0.8$ 
(consistent with the experimental results) and highly oblate for 
$\Omega=0.95$, at which $\lambda'=1.8$.

\begin{figure}[tb]
\psfig{figure=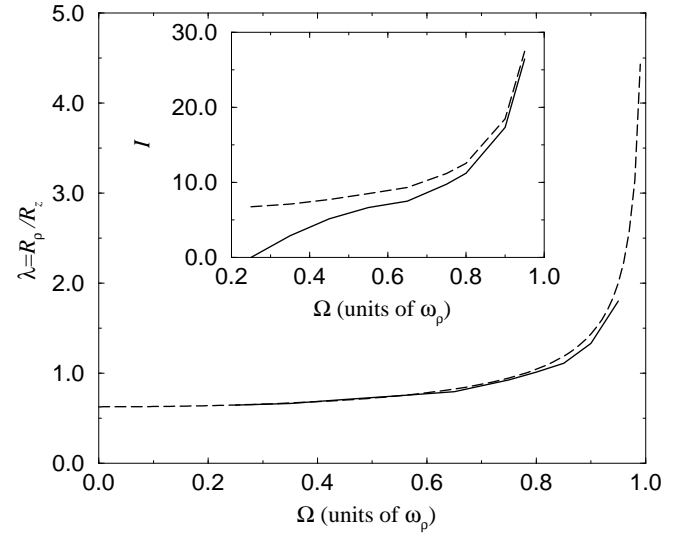,width=\columnwidth,angle=270}
\caption{The cloud aspect ratio $\lambda=R_{\rho}/R_z$ and moment of inertia
$I$ (inset) are shown as a function of rotation frequency $\Omega$. The
numerical results (solid lines) for $\lambda$ are obtained by a TF fit to
the cloud profile, and $I\equiv\langle L_z\rangle/N\Omega$. The solid-body
(dashed lines) result for the cloud anisotropy corresponds to $\lambda
/\sqrt{1-\Omega^2}$, while the corresponding moment of inertia is 
given by $I=\langle x^2+y^2\rangle$.}
\label{sbfig}
\end{figure}

\begin{figure}[tb]
\centering
\subfigure{\psfig{figure=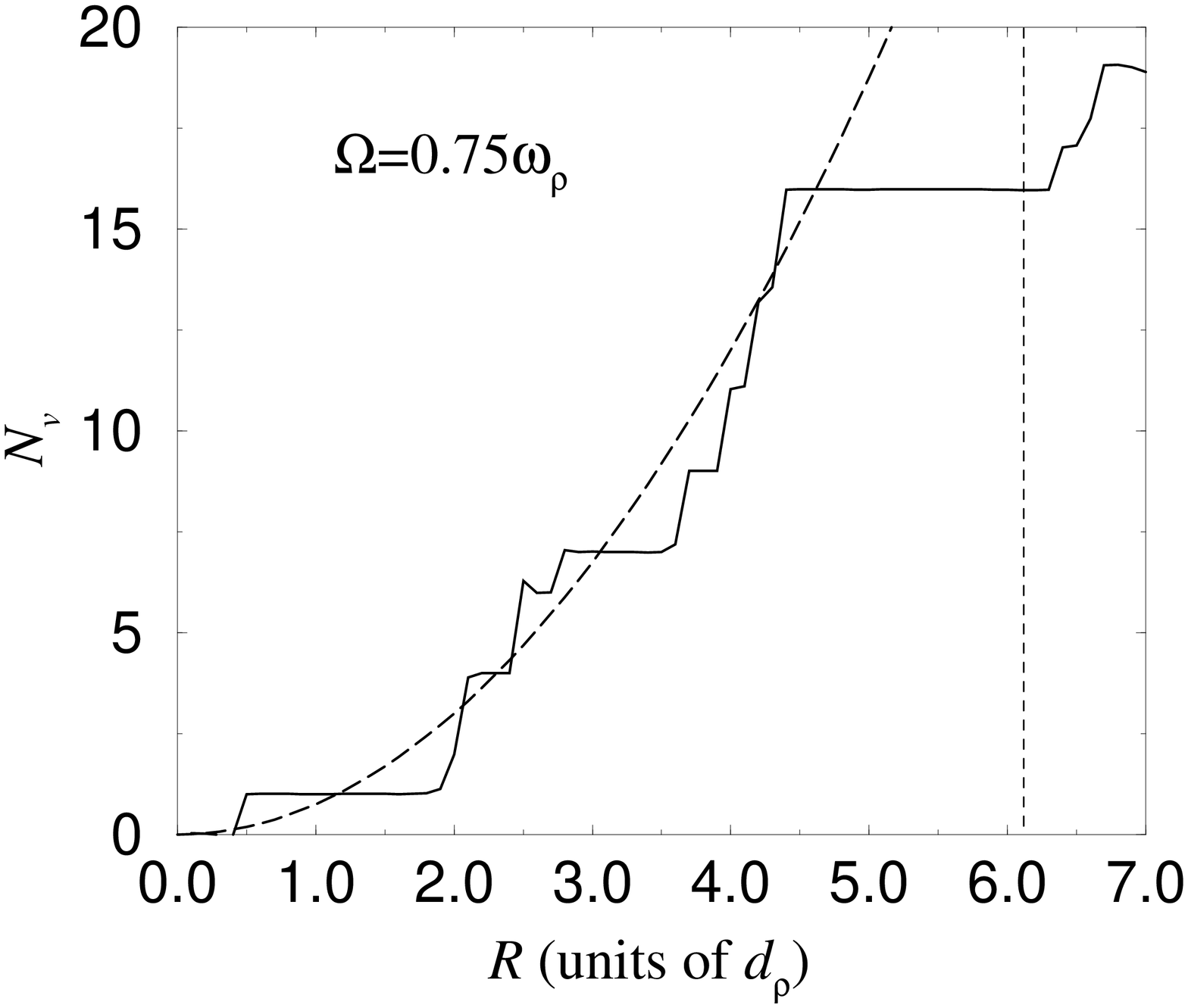,width=0.49\columnwidth,angle=270}}
\subfigure{\psfig{figure=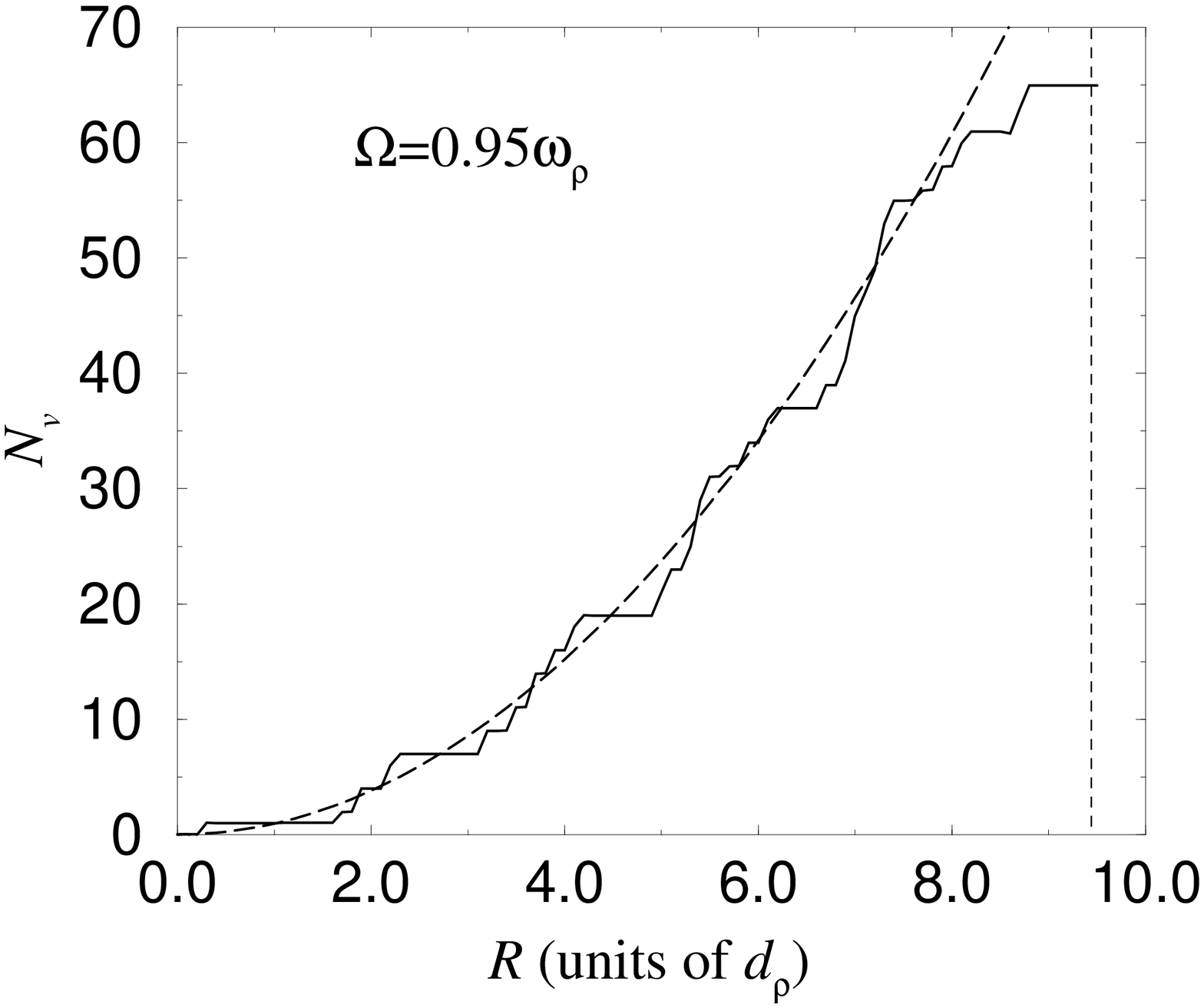,width=0.49\columnwidth,angle=270}}
\caption{The number of vortices within a circular contour centered at the 
origin are shown as a function of radius $R$ (solid lines) for $\Omega=0.75$
and $0.95$. The solid-body predictions $\Omega R^2$ (dashed lines) are shown
for comparison. The vertical dotted lines denote the TF fit for the radial
radius.}
\label{Nvfig}
\end{figure}

As shown in Fig.~\ref{sbfig}, the solid-body estimate of the cloud
anisotropy $\lambda_{\rm sb}'$ tracks (but consistently exceeds) our
numerical values; in contrast, $\lambda_{\rm TF}'$ is always too small. For
example, when $\Omega=0.95$ one obtains $\lambda_{\rm sb}'=2.00$ and
$\lambda_{\rm TF}'=1.50$ with $N_v=65$ (see Fig.~\ref{Nvfig}). The number of
enclosed vortices is not known {\it a priori}, however; using the solid-body
estimate $N_v^{\rm sb}=89$ for $\Omega=0.95$ yields the much improved
$\lambda_{\rm TF}'=1.83$. Another indication that the condensate is behaving
classically at large $\Omega$ is the moment of inertia $I$ (inset of
Fig.~\ref{sbfig}). The effective value $I=\langle L_z\rangle/\Omega$ is always
lower than the solid-body $I=\langle x^2+y^2\rangle$, but is within $4\%$ by
$\Omega=0.95$.

The number of vortices at equilibrium is always considerably lower than the
solid-body prediction, as in previous experimental
observations~\cite{Abo-Shaeer}. Since the numerical solutions are stationary
in the rotating frame, this discrepancy cannot be explained by
positing that the vortex array rotates more slowly than the trap. 
Consider the cases $\Omega=0.55$, $0.8$, $0.9$, and $0.95$ shown
in Fig.~\ref{pix}, which approximate centered triangular arrays 
with $N_v=1+3r(r+1)$, $r=1-4$, respectively. The average vortex spacing is 
found to follow the prediction $b=\sqrt{3/\Omega}$ to within $3\%$.
An additional hexagonal ring of vortices could therefore fit comfortably
within the cloud. For $\Omega=0.95$, $5b=8.89$ is smaller than the
radius $R_{\rho}=9.41$, and $r=5$ corresponds to $N_v=91$ which
is close to the solid-body prediction $N_v^{\rm sb}=89$. For $N_v=169$
($r=7$), which is comparable to the largest array in experiments at
MIT~\cite{Abo-Shaeer}, the missing $n_r=8$ ring implies that the equilibrium
number of vortices is of order $20\%$ lower than the solid-body prediction.

The absence of the last ring might be due to the fact that vortices in this
low-density region would significantly overlap because of their large core 
size. Assuming that the vortex diameter is twice the local healing length, 
then with $\xi(\rho,z=0)=1/R_{\rho}\sqrt{(1-\Omega^2)(1-\rho^2/R_{\rho}^2)}$
one obtains a critical vortex displacement $\rho_c\sim 9$ for $\Omega=0.95$.
In fact, the energy of a uniform array of vortices in a rotating cylinder is
also minimized if there exists a `vortex-free strip' the size of approximately
one ring near the edge of the vessel~\cite{Hall}, i.e.\
$N_v=2\pi R^2\Omega/\kappa-\delta$, where $\delta\sim N_v^{-1}$. This
correction is due to the contribution to the energy of strictly irrotational 
flow in the region between the last vortex and the superfluid surface.

\begin{figure}[tb]
\centering
\subfigure{\psfig{figure=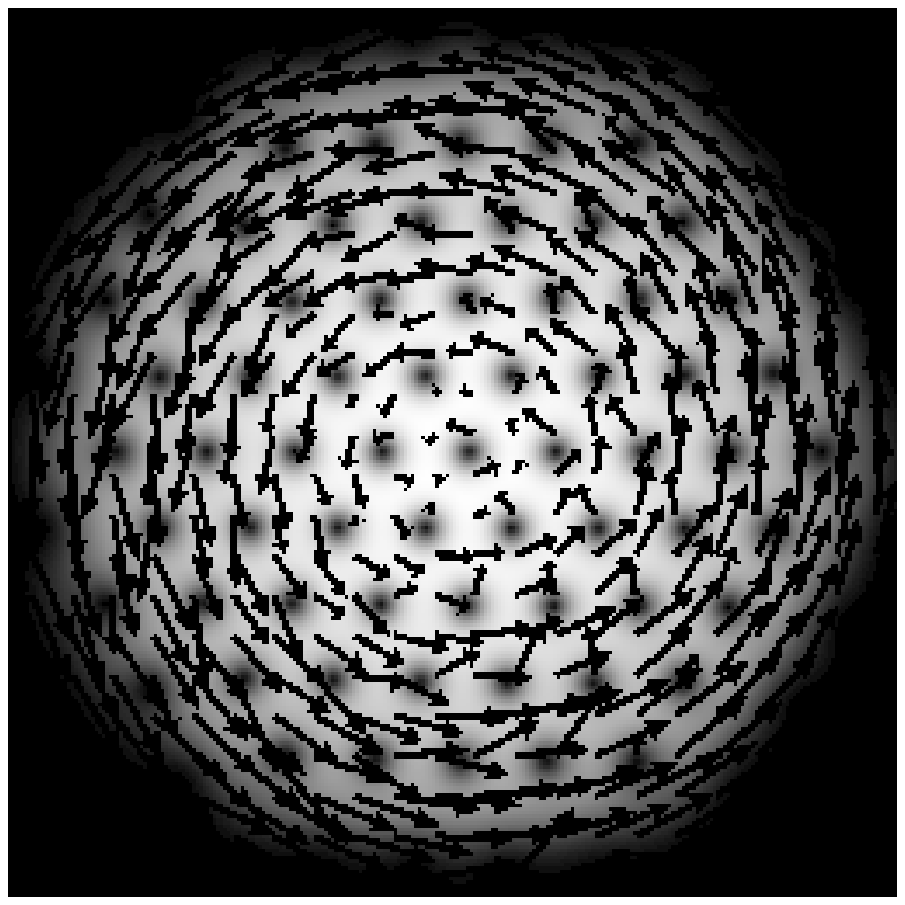,width=0.49\columnwidth,angle=0}}
\subfigure{\psfig{figure=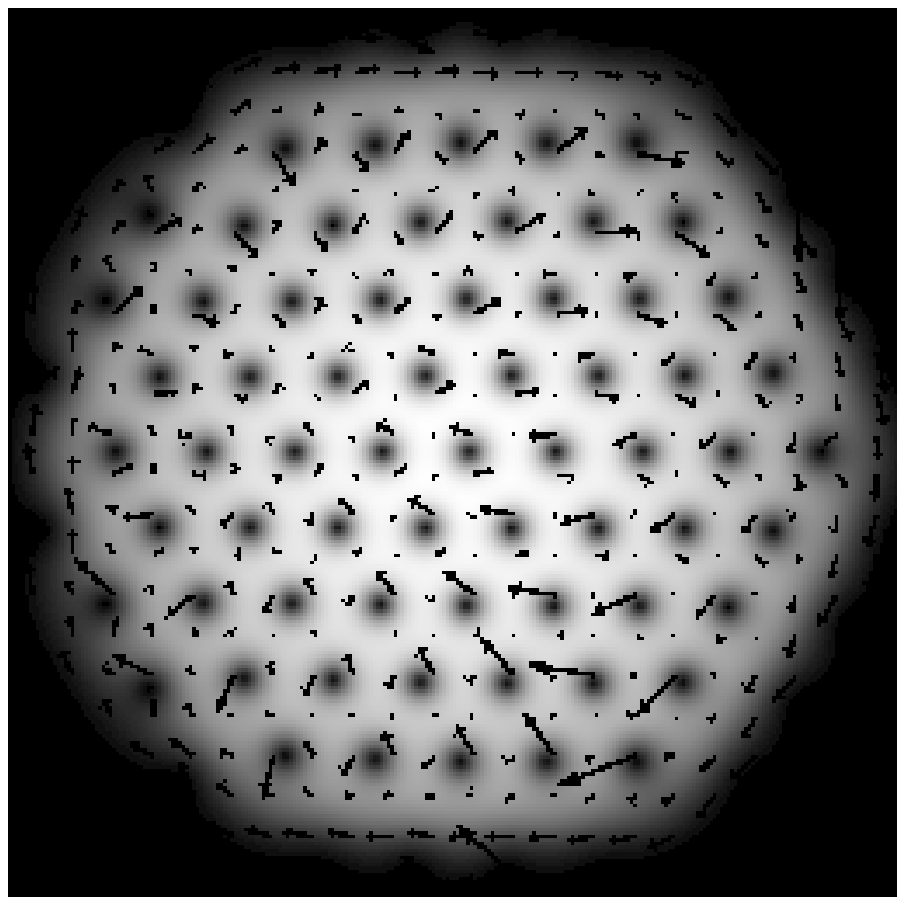,width=0.49\columnwidth,angle=0}}
\caption{The velocity field ${\bf v}$ in the $xy$-plane is represented by
arrows for the $\Omega=0.95$ case. The left and right images correspond to
the lab and rotating frames, respectively.}
\label{velocityfig}
\end{figure}

The existence of a vortex-free region in trapped condensates is confirmed by
evaluating the change in condensate phase around a contour in the $xy$-plane
of increasing radius $R$ from the origin. This is accomplished by
calculating
the spatial derivatives of the numerical data in order to determine
${\bf v}\equiv\nabla\Phi$, interpolating the results onto a one-dimensional
azimuthal grid with 2000 points, and evaluating the line integral
$\oint{\bf v}\cdot d{\bf l}$ numerically using a trapezoidal rule. The
results
for $\Omega=0.75$ and $0.95$ are given in Fig.~\ref{Nvfig}. On average, the
number of vortices follows the solid-body expression $\Omega R^2$ for small
rings, but begins to lag noticeably as $R\to R_{\rho}$ even before the
vortex-free strip is reached. The velocity field for the $\Omega=0.95$ case,
shown in Fig.~\ref{velocityfig}, is small in the rotating frame everywhere 
except for the rotational currents near the vortex cores and the
irrotational 
flow near the surface.

In order to further explore this issue, consider a model wavefunction with
constant amplitude and phase given by $\Phi(x,y)=\sum_{x_0,y_0}\tan^{-1}
[(y-y_0)/(x-x_0)]$, where $(x_0,y_0)$ are vortex positions in a centered
triangular array with lattice constant $b$. For $N_v=61$ ($r=4$), the vortex
velocities $v=|\nabla\Phi|$ on successive hexagonal rings $n_r$ are
$v={1\over b}\{3.63,\,7.23,\,10.69,\,13.57\}$. Since $v(n_r=4)<4v(n_r=1)$ by
$7\%$, the angular velocity of the last ring cannot attain the solid-body
value for any choice of $b$. For large arrays, this mismatch in velocities
varies as $(R/R_{\rho})^5$, and is why significant distortion of the vortex
array from triangular is expected near the superfluid
surface~\cite{Campbell}.

The question that immediately arises is: why are the vortex arrays observed
in confined condensates so perfectly triangular, even very near the surface?
One possible explanation is that a displaced vortex will precess around the 
origin even in the absence of other vortices, due to the inhomogeneous 
external potential. Neglecting vortex curvature (which from Fig.~\ref{pix}
is evidently negligible at large $\Omega$), the additional contribution to
the velocity is $v=[R/(R_{\rho}^2-R^2)]\ln(\xi/R_{\rho})$ in the TF 
limit~\cite{Fetter}. Let us return to the case considered above, with $r=4$, 
and choose $\Omega=0.95$ for 
concreteness. Assuming $R_{\rho}=R_0/(1-\Omega^2)^{3/10}$ and imposing 
$3.63/b+v(R=b)\equiv\Omega b$, one obtains $b=1.98$ and
$v=\{1.88,\,3.76,\,5.62,\,7.37\}$. Thus, including the effect of precession,
the solid-body value $v=4\times 1.88$ at $R=b$ now exceeds the velocity of
the last ring $R=4b$ by only $2\%$.

In conclusion, we have explored the crossover of a confined Bose-Einstein
condensate from that of an irrotational superfluid to a solid body with 
increasing rotation. The external potential is shown to strongly influence
the density and arrangement of the resulting vortices. Many related issues
remain unresolved, however, among them the spin-up of the superfluid by the 
thermal cloud, the upper critical frequency, and the approach to a quantum 
Hall state; these will be the subject of future work.

\begin{acknowledgments}
We are grateful to E.~A.~Cornell and P.~C.~Haljan for numerous fruitful 
discussions. This work was supported by the U.S.\ Office of Naval Research.
\end{acknowledgments}


\begin{references}

\bibitem{Osborne}
D.~V.~Osborne, Proc. Phys. Soc. London~{\bf A63}, 909 (1950).

\bibitem{Khalatnikov}
I.~M.~Khalatnikov, {\it An Introduction to the Theory of Superfluidity}
(W.~A.~Benjamin, New York, 1965).

\bibitem{Donnelly}
R.~J.~Donnelly, {\it Quantized Vortices in Helium II} (Cambridge University
Press, Cambridge, 1991).

\bibitem{Haljan}
P.~C.~Haljan, I.~Coddington, P.~Engels, and E.~A.~Cornell, e-print:
cond-mat/0106362.

\bibitem{Marago}
O.~M.~Marago {\it et al.}, Phys. Rev. Lett.~{\bf 84}, 2056 (2000).

\bibitem{Onofrio}
R.~Onofrio {\it et al.}, Phys. Rev. Lett.~{\bf 85}, 2228 (2000); C.~Raman
{\it et al.}, J.~Low Temp. Phys.~{\bf 122}, 99 (2001).

\bibitem{Fetter}
A.~L.~Fetter and A.~A.~Svidzinsky, J. Phys. Cond. Mat.~{\bf 13}, R135
(2001).

\bibitem{Madison}
K.~W.~Madison, F.~Chevy, W.~Wohlleben, and J.~Dalibard, Phys. Rev.
Lett.~{\bf 84}, 806 (2000).

\bibitem{Abo-Shaeer}
J.~R.~Abo-Shaeer, C.~Raman, J.~M.~Vogels, and W.~Ketterle, Science~{\bf
292},
476 (2001); C.~Raman {\it et al.}, e-print: cond-mat/0106235.

\bibitem{Hodby}
E.~Hodby {\it et al.}, e-print: cond-mat/0106262.

\bibitem{Campbell}
L.~J.~Campbell and R.~M.~Ziff, Phys. Rev. B~{\bf 20}, 1886 (1979).

\bibitem{GP}
E. P. Gross, Nuovo Cimento~{\bf 20}, 454 (1961); L. P. Pitaevskii, Zh. Eksp.
Teor. Fiz.~{\bf 40}, 646 (1961) [Sov. Phys. JETP~{\bf 13}, 451 (1961)].

\bibitem{Eite}
P.~S.~Julienne, F.~H.~Mies, E.~Tiesinga, and C.~J.~Williams, Phys. Rev.
Lett.~{\bf 78}, 1880 (1997).

\bibitem{Tkachenko}
V. K. Tkachenko, Sov. Phys. JETP~{\bf 22}, 1282 (1966).

\bibitem{Ho}
T.-L. Ho, e-print: cond-mat/0104522.

\bibitem{Feder}
B.~I.~Schneider and D.~L.~Feder, Phys. Rev. A~{\bf 59}, 2232 (1999).

\bibitem{Hall}
H.~E.~Hall and W.~F.~Vinen, Phys. Roy. Soc.~{\bf A238}, 215 (1956).

\end{references}
\end{document}